\documentclass[12pt]{article}
\usepackage{graphicx}
\usepackage{amssymb}
\usepackage{amsmath}

\newcommand{\lsim}{ \mathop{}_{\textstyle \sim}^{\textstyle <} }
\newcommand{\bear}{\begin{array}}  
\newcommand {\eear}{\end{array}}
\newcommand{\bea}{\begin{eqnarray}}   
\newcommand{\eea}{\end{eqnarray}}
\newcommand{\beq}{\begin{eqnarray}}   
\newcommand{\eeq}{\end{eqnarray}}
\newcommand{\bef}{\begin{figure}}  \newcommand 
{\eef}{\end{figure}}
\newcommand{\bec}{\begin{center}}  \newcommand 
{\eec}{\end{center}}

\newcommand{\Slash}[1]{{\ooalign{\hfil/\hfil\crcr$#1$}}}

\begin{document}

\begin{titlepage}

\begin{flushright}
IPMU 11-0174~\\
CALT 68-2850\\
\end{flushright}

\vskip 1.35cm
\begin{center}

  {\large {\bf Direct Detection of Dark Matter \\ Degenerate with
      Colored Particles in Mass } }

\vskip 1.2cm

Junji Hisano$^{a,b}$,
Koji Ishiwata$^c$, 
and
Natsumi Nagata$^{a,d}$\\

\vskip 0.4cm
{\it $^a$Department of Physics,
Nagoya University, Nagoya 464-8602, Japan}\\
{\it $^b$Institute for the Physics and Mathematics of the Universe,
University of Tokyo, Kashiwa 277-8568, Japan}\\
{\it $^c$California Institute of Technology, Pasadena, CA 91125, USA}\\
{\it $^d$Department of Physics, 
University of Tokyo, Tokyo 113-0033, Japan}
\date{\today}

\begin{abstract} 
  In this Letter we explore the direct detection of the dark matter in
  the universe, assuming the dark matter particles are degenerate in
  mass with new colored particles below TeV scale.  The scenario with
  such a mass spectrum is difficult to be confirmed or excluded by the
  present analysis at the LHC experiments because the QCD jets in the
  cascade decay of the new particles produced in the proton-proton
  collision are too soft to be triggered in the event selection.  It
  is shown that both of the spin-independent and spin-dependent
  couplings of the dark matter with a nucleon are enhanced and the
  scattering cross section may reach even the current bound of the
  direct detection experiments. Then such a degenerate scenario may be
  tested in the direct detection experiments.
\end{abstract}

\end{center}
\end{titlepage}

%%%%%%%%%%%%%%%%%%%%%%%%%%%%%%%%%%%
\section{Introduction}
%%%%%%%%%%%%%%%%%%%%%%%%%%%%%%%%%%%

A variety of observational evidences indicate the existence of
non-baryonic dark matter (DM).  The latest cosmological observations
including the WMAP reveal that dark matter accounts for about 83\%
of the energy density of matter in the
universe~\cite{Komatsu:2010fb}. The standard model (SM) in
particle physics, however, does not explain its existence, and its
identification has been a challenge in both particle physics and
cosmology. There are several candidates proposed for the dark matter
in physics beyond the SM.  One of the most attractive candidates is
so-called Weakly Interacting Massive Particles (WIMPs), which interact
with ordinary matter only through weak and gravitational
interactions. In the WIMP DM scenario, WIMPs are supposed to be stable
and produced in the thermal history of the universe. Then naturally
their relic explains the present abundance of dark matter with the
mass around TeV scale.

Nowadays we are reaching to the energy frontier of the TeV scale at
the collider experiments. At the Large Hadron Collider (LHC), where
protons are colliding at the center of mass energy of 7 TeV, the WIMP
dark matter is expected to be produced and observed as large missing
energy.  In addition, several extensions of the SM also include heavy
colored particles which interact with WIMPs directly or via gauge
bosons.  Once they are produced at the collider, they subsequently
decay into WIMPs accompanied by high energy QCD jets. That is why the
main strategy for probing such new models is based on events with hard
jets and large missing transverse energy. The LHC experiments have
already presented results for several models, and put severe
constraints on them as no excess above the SM background has been
observed. For example, the exclusion limits for the Constrained
Minimal Supersymmetric Standard Model (CMSSM) are presented in
Refs.~\cite{Aad:2011ib, Chatrchyan:2011zy}. However, the constraints
are not directly applied if the colored particles are nearly
degenerate with the WIMPs in mass~\cite{Kawagoe:2006sm}. In this case,
the leading jets in most events are too soft to pass the signal
selection, and thus it is difficult to confirm or exclude such
scenario at the LHC.

In this Letter, we study possibilities for probing this scenario in
terms of the direct detection experiments of the dark matter in the universe.
We found that both of the spin-independent (SI) and spin-dependent (SD)
couplings of the WIMPs with a nucleon are enhanced when the masses of
colored particles which mediate the scattering are degenerate with
that of WIMPs.  This implies that DM direct detection experiments
offer a promising way to investigate the degenerate scenario.  

The ongoing direct detection experiments have extremely high
sensitivities. For example, XENON100 currently gives the most
stringent constraint on the SI WIMP DM-nucleon elastic scattering
cross sections, $\sigma_{\rm SI}<7.0\times 10^{-45}~{\rm cm}^2$ for a
WIMP mass of 50 GeV \cite{Aprile:2011hi}. Moreover, there are several
proposals to use ton-scale detectors in the future, which are designed
to detect WIMPs with sensitivities much below the current limit.
Recently a proposal for a future WIMP search experiment, which is
planning to achieve a sensitivity of the SI cross section,
$10^{-49}~{\rm cm}^2$, is announced~\cite{Malling:2011va}.  With
regard to the SD cross section, the IceCube detector would offer
excellent sensitivities to it~\cite{Helbing:2011wf}. It has provided
upper limits on the SD cross section of WIMPs with a nucleon,
$\sigma_{\rm SD}<10^{-(39-40)}$~cm$^{2}$ with WIMP mass in the
range of $100~{\rm GeV}$--$1~{\rm TeV}$, and further improvement of
the sensitivities by a factor of 2--10 is expected in the same
mass range~\cite{TeVPAtalk, Abbasi:2009vg}.

The scattering cross section of WIMPs with a nucleon is evaluated by
using the several effective couplings of WIMPs with quarks and
gluon. The precise computation of them is important for the evaluation
of the cross section because they may be either destructive
\cite{Hisano:2010fy,Hisano:2011cs} or constructive
\cite{Hisano:2010yh}, depending on WIMP models. Recently an approach
to study nature of the dark matter based on both collider and direct
detection experiments in terms of the effective theory is proposed
\cite{Goodman:2010ku,Fox:2011pm}. The authors, however, take just one
of the effective couplings into account for the evaluation of
the cross sections. In this Letter, we include all of the relevant
effective operators into the calculation, and find that the scattering
amplitude is enhanced when the masses of colored particles which
mediate the scattering are degenerate with that of WIMPs.

In our Letter we study two representative models; the minimal
supersymmetric standard model (MSSM) and the minimal universal extra
dimension model (MUED).  In the MSSM, the lightest superparticle
(LSP), which is stabilized by the $R$ parity, is a good candidate for
the DM. In this framework, we consider the lightest neutralino,
especially Wino-like, DM. Among the components of neutralino, Wino has
relatively large coupling to colored particles ({\it i.e.} squarks)
and has a big impact on the cross section. In the MUED model, the
Kaluza-Klein (KK) parity protects the lightest KK particle (LKP) from
decay, and the LKP is a good candidate for the DM. We consider the
scenario where the first excited KK gauge boson is the
LKP~\cite{Servant:2002aq, Cheng:2002ej}.  
There is an earlier literature which studies the complementarity of
direct detection and collider searches in the MUED~\cite{Arrenberg:2008wy}. 
In their work, however, several effective operators are not considered;
thus, the resultant cross sections are smaller than those obtained in
Ref.~\cite{Hisano:2010yh}, where all of the relevant effective operators
are taken into account. We find that the scattering
cross sections may reach even the current bound of the direct
detection experiments.  Although the scattering cross section depends
on the details of DM scenarios, our results obtained for the two
scenarios describe general aspects of enhancement of cross section in
the degenerate scenario.

%%%%%%%%%%%%%%%%%%%%%%%%%%%%%%%%
\section{Wino dark matter}
%%%%%%%%%%%%%%%%%%%%%%%%%%%%%%%%
Let us study the Wino-like neutralino DM case for a starter. To make
discussion simple, we assume the lightest neutralino is close to a pure
Wino state.  Generalization to other composition of the lightest
neutralino is given later.

First of all we give the formulation of the scattering cross section.
Wino is a Majorana fermion, and the phenomena of such a fermion
scattered by nucleon is described in terms of the effective Lagrangian
given in Refs.~\cite{Drees:1993bu,Hisano:2010ct},
\begin{eqnarray}
{\cal L}_{\rm eff}&=&\sum_{q}
\bigl(f_q m_q\ \bar{\tilde{W}}^0  \tilde{W}^0\bar{q}q 
+d_q \bar{\tilde{W}}^0 \gamma_\mu\gamma_5 \tilde{W}^0\bar{q}
\gamma^\mu\gamma_5q \nonumber\\ 
&&+ \frac{g^{(1)}_q}{M}  \bar{\tilde{W}}^0 i \partial^{\mu}\gamma^{\nu} 
\tilde{W}^0  {\cal O}_{\mu\nu}^q
+ \frac{g^{(2)}_q}{M^2}
\bar{\tilde{W}}^0  (i \partial^{\mu})(i \partial^{\nu})
\tilde{W}^0  {\cal O}_{\mu\nu}^q 
\bigr) \nonumber \\ 
&&+f_G \bar{\tilde{W}}^0 \tilde{W}^0 G_{\mu\nu}^aG^{a\mu\nu} \, .
\label{LeffWino}
\end{eqnarray}
Here $\tilde{W}^0$ and $q$ denote Wino and light quark with masses $M$
and $m_q$, respectively, and $G^a_{\mu\nu}$ is the field strength
tensor of gluon field.  The twist-2 operator, ${\cal O}_{\mu\nu}^q$,
is defined as $ {\cal O}_{\mu\nu}^q\equiv\frac12 \bar{q} i
\left(D_{\mu}\gamma_{\nu} + D_{\nu}\gamma_{\mu}
  -\frac{1}{2}g_{\mu\nu}\Slash{D} \right) q$ with the covariant
derivative $D_{\mu}$. Then the elastic scattering cross section of DM
with nucleon ($N=p,n$) is obtained from the effective Lagrangian as
\begin{eqnarray}
  \sigma_{\tilde{W}^0N}=
  \frac{4}{\pi}m_r^2
  \left[\left| f_N\right|^2+3\left|a_N\right|^2\right]\, , 
\label{sigma}
\end{eqnarray}
where $m_r= M m_N/(M+m_N)$ is the reduced mass with $m_N$ being the
nucleon mass. The SI effective coupling, $f_N$, in Eq.~(\ref{sigma})
is evaluated by using the nucleon matrix elements of the quark and
gluon operators in the effective Lagrangian. The result is
\begin{eqnarray}
\frac{f_N}{m_N}=\sum_{q=u,d,s}
f_q f_{Tq}
+\sum_{q=u,d,s,c,b}
\frac{3}{4} \left(q(2)+\bar{q}(2)\right)\left(g_q^{(1)}+g_q^{(2)}\right)
-\frac{8\pi}{9\alpha_s}f_{TG} f_G \, ,
\label{f}
\end{eqnarray}
where 
$\langle N \vert m_q \bar{q} q \vert N\rangle/m_N = f_{Tq}$, 
$f_{TG}= 1-\sum_{u,d,s}f_{Tq}$  and $\langle N(p)\vert 
{\cal O}_{\mu\nu}^q
\vert N(p) \rangle 
=\frac{1}{m_N}
(p_{\mu}p_{\nu}-\frac{1}{4}m^2_N g_{\mu\nu})
(q(2)+\bar{q}(2))$.
Here $\alpha_s\equiv g_s^2/4\pi$ is the strong coupling
  constant, and $q(2)$ and $\bar{q}(2)$ indicate the second moments
of the parton distribution functions (PDFs).  The SD effective
coupling, on the other hand, is given as
\begin{equation}
 a_{N}=\sum_{q=u,d,s} d_q \Delta q_N \ ,
\label{an}
\end{equation}
where$ \langle N \vert \bar{q}\gamma_{\mu}\gamma_5 q \vert N \rangle
\equiv 2 s_{\mu}\Delta q_N$ with $s_{\mu}$ the spin of the nucleon.
In this Letter we use the values of these matrix elements given in,
{\it e.g.}, Ref.~\cite{Hisano:2010ct}, in which they are evaluated
based on works in
Refs.~\cite{Ohki:2008ff,Cheng:1988im,Pumplin:2002vw}.

Our remaining task is to calculate the coefficients of the effective
operators in Eq.~(\ref{LeffWino}). Wino interacts with quarks and
gluon via the two types of interactions; the Wino-quark-squark
interaction and the weak interaction. The Lagrangian is
\begin{eqnarray}
{\cal L}_{\tilde{W}^0}=\sum_{q}\sum_{i=1,2}
\bar{q} (a_{\tilde{q}_i}+b_{\tilde{q}_i} \gamma_5) 
\tilde{W}^0 \tilde{q}_i
 -g_2(\bar{\tilde{W}}^0\gamma^{\mu}\tilde{W}^-
W^{\dagger}_{\mu}) + {\rm h.c.}\, ,
\label{Lwino}
\end{eqnarray}
where $\tilde{W}^-$ and $W_\mu$ represent the charged Wino and the $W$
boson, respectively, and $\tilde{q}_i~(i=1,2)$ denote the squark mass
eigenstates.  $g_2$ is the $SU(2)_L$ gauge coupling constant. The
Wino-quark-squark interaction induces tree-level scattering, which
gives dominant contribution to the cross section in our present
degenerate scenario, as we will see soon. The coefficients
$a_{\tilde{q}_i}$ and $b_{\tilde{q}_i}$ depend on the squark mixing
angle.  The squark mass term is given as
\begin{equation}
 {\cal L}_{{\rm mass}}=-
\begin{pmatrix}
\tilde{q}_L^* &\tilde{q}_R^*
\end{pmatrix}
\begin{pmatrix}
 m^2_{\tilde{q}_L} & m^2_{\tilde{q}_{LR}} \\
 m^{2*}_{\tilde{q}_{LR}} & m^2_{\tilde{q}_R}
\end{pmatrix}
\begin{pmatrix}
 \tilde{q}_L \\ \tilde{q}_R
\end{pmatrix}\, ,
\label{squark_mass_matrix}
\end{equation}
where $\tilde{q}_L$ and $\tilde{q}_R$ represent the left-handed and
right-handed squarks, respectively, and each component is
\begin{eqnarray}
 m^2_{\tilde{q}_L,\tilde{q}_R}&=&\tilde{m}^2_{\tilde{q}_L,\tilde{q}_R}
+m_q^2+D_{\tilde{q}_L,\tilde{q}_R}\, ,
\\
m^2_{\tilde{u}_{LR}}&=&m_u(A_u^*-\mu\cot\beta) \, , \nonumber \\
m^2_{\tilde{d}_{LR}}&=&m_d(A_d^*-\mu\tan\beta)\, .
\end{eqnarray}
Here $m_u$ and $m_d$ indicate up-type and down-type quark
masses, and $\tilde{m}^2_{\tilde{q}_L,\tilde{q}_R}$ and $A_q$ are
for the soft supersymmetry breaking parameters. $\mu$ denotes the
supersymmetric mass term of the Higgs superfields. (As we mentioned at
the beginning, we consider the pure Wino case. So we take $\mu$ much
larger than Wino mass in the following numerical calculation.)
$\tan\beta$ is the fraction of vacuum expectation values of the
up-type and down-type Higgs fields. $D_{\tilde{q}_L,\tilde{q}_R}$ is
$D$-term contribution given as $D_{\tilde{q}_L}=m_Z^2
\cos2\beta(T^3_{q}-Q\sin^2\theta_W)$ and $D_{\tilde{q}_R}=m_Z^2
\cos2\beta Q\sin^2\theta_W$. ($m_Z$, $\theta_W$, $T^3_{q}$ and $Q$ are
the $Z$ boson mass, the weak-mixing angle, $SU(2)_L$ and the electric
charge of squark, respectively.)  By diagonalizing the mass matrix, we
obtain $a_{\tilde{q}_1}=b_{\tilde{q}_1}$$=-g_2T^3_{qL}\cos
\theta_q/\sqrt{2}$ and
$a_{\tilde{q}_2}=b_{\tilde{q}_2}=g_2T^3_{qL}\sin
\theta_q/\sqrt{2}$. Here $\tilde{q}_1$ is the lighter squark and
$\sin 2\theta_q\equiv
2m^2_{{\tilde{q}_{LR}}}/(m^2_{\tilde{q}_1}-m^2_{\tilde{q}_2})$.

With those expression for $a_{\tilde{q}_i}$ and $b_{\tilde{q}_i}$, 
the coefficients in the effective coupling in
Eq.~(\ref{f}) induced by squark exchange at tree
level are derived as follows~\cite{Drees:1993bu,Hisano:2010ct},
\begin{eqnarray}
f_q&=&\frac{g_2^2M}{32}\biggl[
\frac{\cos^2\theta_q}{(m_{\tilde{q}_1}^2-M^2)^2}
+\frac{\sin^2\theta_q}{(m_{\tilde{q}_2}^2-M^2)^2}
\biggr]
\, ,\nonumber\\
d_q&=&\frac{g_2^2}{16}\biggl[
\frac{\cos^2\theta_q}{m_{\tilde{q}_1}^2-M^2}
+\frac{\sin^2\theta_q}{m_{\tilde{q}_2}^2-M^2}
\biggr]
\, ,\nonumber\\
g_q^{(1)}&=&\frac{g_2^2M}{8}\biggl[
\frac{\cos^2\theta_q}{(m_{\tilde{q}_1}^2-M^2)^2}
+\frac{\sin^2\theta_q}{(m_{\tilde{q}_2}^2-M^2)^2}
\biggr]
\, ,\nonumber\\
g_q^{(2)}&=&0 \, .
\label{effective_couplings} 
\end{eqnarray}
It is clear in the above expression that they are enhanced when the
squark masses are degenerate with the Wino mass. The Wino-squark-quark
interaction also induces scattering with gluon at one loop. Although
this contribution is $O(\alpha_s)$, it is comparable to the tree-level
ones in the SI effective coupling $f_N$ in Eq.~(\ref{f}) since the term
proportional to the effective scalar coupling of gluon, $f_G$, has a
factor of $1/\alpha_s$.
According to Ref.~\cite{Hisano:2010ct}, the gluon
contribution is evaluated as
\begin{equation}
 f_G=-\frac{\alpha_s g_2^2M}{384\pi}\biggl[
\frac{\cos^2\theta_q}{m_{\tilde{q}_1}^2(m_{\tilde{q}_1}^2-M^2)}
+\frac{\sin^2\theta_q}{m_{\tilde{q}_2}^2(m_{\tilde{q}_2}^2-M^2)}
\biggr]~.
\end{equation}
This result is for the case where only first generation squarks are
degenerate with dark matter in mass. We will focus on such a scenario in
our numerical calculation, which will be addressed soon. From its
explicit expression, it is easily seen that the gluon contribution is
also enhanced in the degenerate mass spectrum.

In addition to the squark exchanging process, Wino interacts with
quarks and gluons through the weak interaction at loop level.  These
contributions might be sizable especially in the case where Wino is
much heavier than the weak scale since they are not suppressed even in
such a spectrum~\cite{Hisano:2004pv}. All of these contributions are
evaluated in Ref.~\cite{Hisano:2010fy, Hisano:2010ct}, and we include
them into our numerical calculation.

The phenomena at the LHC in the supersymmetric model is followed by
pair production of colored superparticles: $pp\rightarrow
\tilde{q}\tilde{g}$, $\tilde{q}\tilde{q}$ and $\tilde{g}\tilde{g}$
where $\tilde{g}$ is gluino and $\tilde{q}$ is squarks (mainly
  the first generation squarks).  Then subsequent cascading decay of
$\tilde{q}$ and $\tilde{g}$ to the LSP yields many QCD jets and
missing energy since the LSPs escape from the detectors without
tracks. At the present stage of the data analysis, the masses of
$\tilde{q}$ and $\tilde{g}$ are constrained nearly up to $1~{\rm TeV}$
due to the null signal event~\cite{Aad:2011ib, Chatrchyan:2011zy}. In
our Letter, we discuss the case where the Wino LSP is so degenerate with
squarks in mass that the missing transverse energy signature would not
be observed at the LHC experiments due to soft jets. Since triggers
for the leading jets in their analysis are typically set to be above
$100$ GeV ({\it e.g.}, the leading jets are required to have the
transverse momentum larger than 130 GeV in ATLAS
  Collaboration~\cite{Aad:2011ib}, while each of the two
hardest jets in events must have the transverse energy larger than 100
GeV in CMS Collaboration~\cite{Chatrchyan:2011zy}), the
degeneracy of 100 GeV in their masses is enough to conceal the missing
energy signals at the present stage of the data analysis at the LHC
experiments.\footnote{Here we note there are several works which
  investigate collider signature in the degenerate mass spectrum
  scenarios by using initial state radiation in the
  MSSM~\cite{Alwall:2008ve,Alwall:2008va} and $M_{T2}$ in the
  MUED~\cite{Murayama:2011hj}.}  Even when the mass difference is
200--300 GeV, it is hard to probe the signature by using the current
approach.\footnote{We thank S. Asai for his instruction in private
  communications. }

For simplicity, we assume
the first generation squarks to be degenerate with Wino and the other
squarks to be heavy enough to evade the current bound. Gluino is
assumed to be either degenerate with Wino or much heavier than the
present limit. The degeneracy is parametrized as
\begin{equation}
\Delta m\equiv \tilde{m}_{\tilde{q}_L,\tilde{q}_R}-M \, .
\end{equation}
Considering  the above discussion, we carry out the calculation
with the parameter $\Delta m$ up to 200 GeV.
%%%%%%%%%%%%%%%%%%FIGURE%%%%%%%%%%%%%%%%%%%
\begin{figure}[t]
 \begin{center}
  \includegraphics[height=8cm]{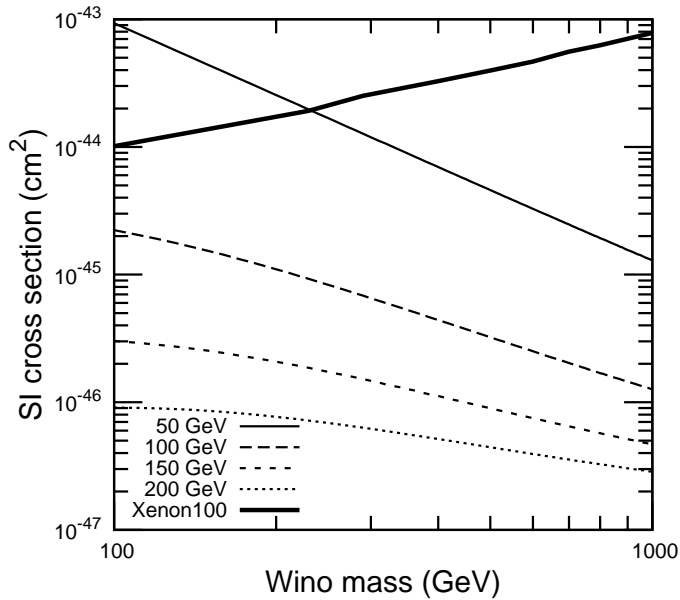}
  \caption{SI scattering cross section of Wino DM with a
    proton as a function of Wino-like neutralino mass. Each line
    corresponds to $\Delta m=50$, $100$, $150$ and
    $200~{\rm GeV}$ from top to bottom, and  
    upper bound from XENON100 ~\cite{Aprile:2011hi} is shown in bold line.}
\label{fig:wino1st}
 \end{center}
\end{figure}
%%%%%%%%%%%%%%%%%%%%%%%%%%%%%%%%%%%

In Fig.~\ref{fig:wino1st}, we plot the SI scattering cross section of
the Wino DM with a proton as a function of the DM mass. Each line
corresponds to the case where $\Delta m=50$, $100$, $150$ and
$200~{\rm GeV}$ from top to bottom. In the figure, the limit by
  the XENON100 experiment~\cite{Aprile:2011hi} is depicted in a bold
  line. Here we take the other parameters as $A_{u,d}=0$,
$\mu=M+1~{\rm TeV}$, $\tan \beta=10$ and $m_h=120~{\rm GeV}$ in order
to include the SM Higgs boson contribution to the SI cross section.
We have checked that the cross section has little dependence on those
parameters. The result given in the figure shows that the SI cross
section is considerably enhanced when squarks are degenerate with Wino
in mass and it is quite sensitive to the degeneracy. In the
calculation for Fig.~\ref{fig:wino1st}, we found that the `twist-2'
contribution with coefficient $g_q^{(1)}$ in Eq.~(\ref{f}) is the main
contribution as expected from Eq.~(\ref{effective_couplings}). When
$\Delta m =50~{\rm GeV}$, the Wino mass of less than $200~{\rm GeV}$
is excluded by the XENON100 result. Even in the case of $\Delta m
=200~{\rm GeV}$ the SI cross section is $10^{-46}$--$10^{-47}~{\rm
  cm}^2$ for $M=100~{\rm GeV}$--$1~{\rm TeV}$. Such a value of the
cross section would be tested by future experiments.

We can also consider the case where other squarks, {\it e.g.,} the
third generation squarks, are degenerate with the lightest neutralino
in mass instead of the first generation squarks. In such cases, the
scattering cross section tends to be rather small because the
tree-level contribution is suppressed. However, in some parameter
region, the SI cross section could be large enough to be accessible in
the future direct detection experiment.

%%%%%%%%%%%%%%%%%%FIGURE%%%%%%%%%%%%%%%%%%%
\begin{figure}[t]
 \begin{center}
  \includegraphics[height=8cm]{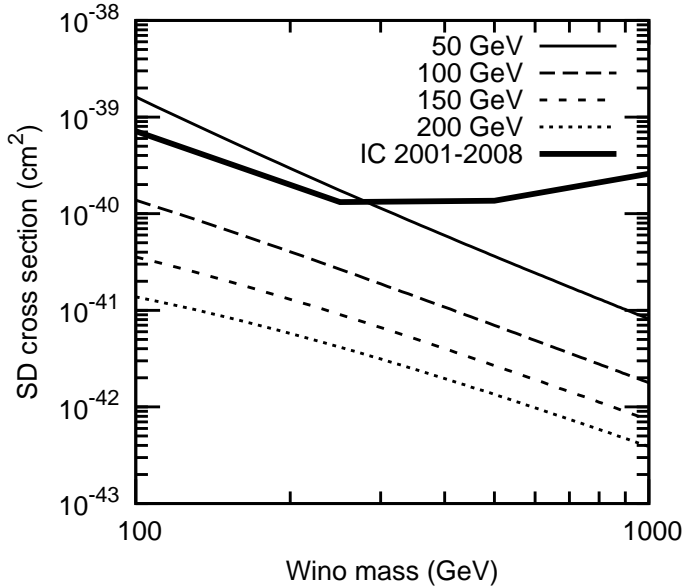}
  \caption{SD scattering cross section of Wino DM with a proton as a
    function of Wino-like neutralino mass. Parameters in
      this plot and line contents are the same as
    Fig.~\ref{fig:wino1st}. Preliminary IceCube
      bound~\cite{TeVPAtalk} is also shown in bold line. This bound is
  given by assuming $WW$ final state, which is suitable for our present
  Wino DM.}
\label{fig:winosd}
 \end{center}
\end{figure}
%%%%%%%%%%%%%%%%%%%%%%%%%%%%%%%%%%%
Next we show the SD scattering cross section of Wino DM with a proton
as a function of Wino mass in Fig.~\ref{fig:winosd}.  In the plot the
parameters are taken to be the same values as those for the SI cross
section evaluated above. We observe the similar enhancement due to the
mass degeneracy of DM with squarks in the SD scattering cross section,
as is expected. When $\Delta m\lesssim 100$~GeV, the SD cross section
is comparable to the sensitivity of IceCube experiment,
$\sigma_{\rm  SD}\lsim  10^{-(40-41)}$~cm$^{2}$ \cite{TeVPAtalk}.

So far we have discussed the pure Wino DM scenario. To end this
section, we give some comments on the extension to more general
neutralino DM.  When $\mu$ is not extremely large compared to the weak
scale, the lightest neutralino is no longer a pure Wino state, rather
the mixed state of Bino, Wino and Higgsinos.  For example, the
Wino-like neutralino interacts with the Higgs bosons (SM-like and
heavy Higgs bosons) and the $Z$ boson via mixing with Higgsinos, which
gives rise to the Higgs boson exchange in the SI scattering and the
$Z$ boson exchange in the SD scattering, respectively. We computed
these types of contributions and found that they might have sizable
effects. When $\mu \lesssim M+400~{\rm GeV}$, the contribution from
the SM-like Higgs boson exchange has the opposite sign of the twist-2
contribution and its absolute value may be comparable to the twist-2
contribution. On the other hand, the heavy Higgs boson contribution
depends on the sign of $\mu$, the heavy Higgs boson mass and $\tan
\beta$. If $\mu$ is positive, this contribution has the same sign as
the twist-2 contribution and vice versa. As to its absolute value, it
is suppressed by the heavy Higgs boson mass, while it is enhanced by
$\tan \beta$.  When $\tan \beta$ is large ({\it e.g.}, $\sim 30$), it
gives a contribution to the effective coupling, almost comparable to
the SM-like Higgs one, despite the large mass of the heavy Higgs boson
around $1~{\rm TeV}$.  Therefore this type of Higgs contribution could
be either constructive or destructive in the SI effective coupling,
depending on the parameters in the Higgs sector in the MSSM.
The $Z$ boson exchange process also might yield a sizable effect to
the SD cross section, however, the squark exchange process tends
to dominate the effective coupling in the degenerate scenario.

As regards Bino-like and Higgsino-like neutralino DM, we have also
evaluated their impact on the scattering cross sections with
nucleon. In the case of Bino-like DM, we found that the cross section
is in general smaller than that of Wino DM. 
For Higgsino-like DM, the cross section is also suppressed due to its
considerably small couplings to the first and second generation
squarks unless the gaugino-Higgsino mixing is sizable.  The details
will be given elsewhere~\cite{HIN}.

%%%%%%%%%%%%%%%%%%%%%%%%%%%%%%%%%
\section{Kaluza-Klein dark matter}
%%%%%%%%%%%%%%%%%%%%%%%%%%%%%%%%%%%

In this section, we consider the case of LKP DM in the framework of
MUED model.  In the MUED model, we consider the first excited state of
the $U(1)_Y$ $B$ boson, $B^{(1)}_{\mu}$, is the LKP and becomes the
dark matter. The effective Lagrangian of such vector dark matter with
quarks and gluon is formulated in Ref.~\cite{Hisano:2010yh}:
\begin{eqnarray}
\mathcal{L}_{\mathrm{eff}}&=&\sum_{q}
\bigl(f_q m_q B^{\mu}B_{\mu}\bar{q}q+\frac{d_q}{M}
  \epsilon_{\mu\nu\rho\sigma}B^{\mu}i\partial^{\nu}B^{\rho}
  \bar{q}\gamma^{\sigma}\gamma^{5}q+
\frac{g_q}{M^2}
  B^{\rho}i\partial^{\mu}i\partial^{\nu}B_{\rho}\mathcal{O}^q_{\mu\nu}
\bigr) \nonumber \\
&&+ f_G B^{\rho}B_{\rho}G^{a\mu\nu}G^a_{\mu\nu}\, ,
 \label{LeffLKP}
\end{eqnarray} 
where $\epsilon^{\mu\nu\rho\sigma}$ is the totally antisymmetric
tensor defined as $\epsilon^{0123}=+1$.  Here we omit the superscript
of $B^{(1)}_{\mu}$ for simple expression. In this case the elastic
scattering cross section of the DM with nucleon is given as
\begin{eqnarray}
  \sigma_{B^{(1)}N}=
  \frac{m^2_r}{\pi M^2} 
  \left[\left| f_N\right|^2+2\left|a_N\right|^2\right]\, .
\label{sigma_vec}
\end{eqnarray}
In this case, the SI effective coupling is given by Eq.~(\ref{f})
  in which $(g^{(1)}_q+g^{(2)}_q)$ is replaced by $g_q$, and the SD effective 
  coupling is expressed the same as given by Eq.~{(\ref{an})}.
All of the coefficients of the effective operators in
Eq.~(\ref{LeffLKP}) are computed in Ref.~\cite{Hisano:2010yh}. The
tree-level contributions are
\begin{eqnarray}
f_q&=&-\frac{g_1^2}{4m_h^2}-\frac{g_1^2}{4}\biggl[Y^2_{\mathrm{qL}}
\frac{m^2_{Q^{(1)}}}{(m^2_{Q^{(1)}}-M^2)^2}+
Y^2_{\mathrm{qR}}\frac{m^2_{q^{(1)}}}{(m^2_{q^{(1)}}-M^2)^2}\biggr] 
 \nonumber \\ &&
+\frac{g_1^2Y_{\mathrm{qL}}Y_{\mathrm{qR}}}{m_{Q^{(1)}}+m_{q^{(1)}}}
\biggl[\frac{m_{Q^{(1)}}}{m^2_{Q^{(1)}}-M^2}
+\frac{m_{q^{(1)}}}{m^2_{q^{(1)}}-M^2}\biggr] \, , 
\label{fq} \\
d_q&=&\frac{ig^2_1M}{2}\biggl[\frac{Y^2_{\mathrm{qL}}}{m^2_{Q^{(1)}}-M^2}
+\frac{Y^2_{\mathrm{qR}}}{m^2_{q^{(1)}}-M^2}\biggr]\, , \\
g_q&=&-g_1^2M^2\biggl[\frac{Y^2_{\mathrm{qL}}}{(m^2_{Q^{(1)}}-M^2)^2}+
\frac{Y^2_{\mathrm{qR}}}{(m^2_{q^{(1)}}-M^2)^2}\biggr]\, .
\end{eqnarray}
Here $q^{(1)}$ and $Q^{(1)}$ describe the mass eigenstates of the
first KK quarks which are $SU(2)_L$ singlet and doublet with masses
$m^2_{q^{(1)}}$ and $m^2_{Q^{(1)}}$, respectively. $Y_{\rm qL}$ and
$Y_{\rm qR}$ are the hypercharges of left-handed and right-handed
quarks, and $g_1$ is the $U(1)_Y$ gauge coupling constant. The first
term in Eq.~(\ref{fq}) corresponds to the SM Higgs boson exchange
contribution, while the other terms come from the KK quark exchange
processes. The KK quark-exchange processes turn out to be enhanced
when their masses are degenerate with the DM mass, similarly to
the previous Wino case. For the gluon contribution, we use the
results in Ref.~\cite{Hisano:2010yh}. To make our discussion simple,
we take mass parameters of the first KK quarks $m_{q1,Q1}$ as
\begin{eqnarray}
m_{q1,Q1} = M+\Delta m 
\end{eqnarray}
to give $m^2_{q^{(1)},Q^{(1)}}=m^2_{q1,Q1}+m^2_q$, and consider the
parameter region $\Delta m\lesssim 200$ GeV. Such mass degeneracy is
generally obtained in the MUED model since the mass of each KK mode
at tree-level is just determined by the radius of the extra dimension.

%%%%%%%%%%%%%%%%%%FIGURE%%%%%%%%%%%%%%%%%%%
\begin{figure}[t]
 \begin{center}
  \includegraphics[height=8cm]{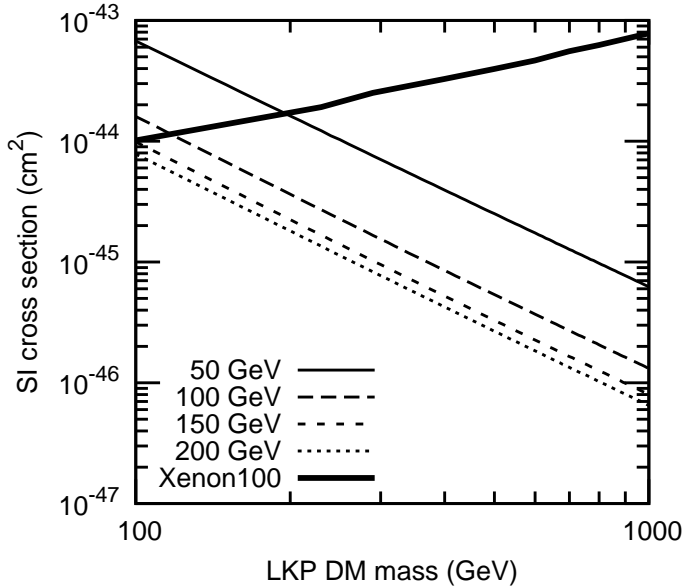}
  \caption{SI scattering cross section of LKP DM with a proton
    as a function of DM mass. From top to bottom
    $\Delta m=50$, $100$, $150$ and $200~{\rm GeV}$, and XENON100 limit
  is also shown in
    bold line~\cite{Aprile:2011hi}.}
\label{fig:lkp}
 \end{center}
\end{figure}
%%%%%%%%%%%%%%%%%%%%%%%%%%%%%%%%%%%
Now we are ready to evaluate the SI cross section of LKP DM with a
proton. Fig.~\ref{fig:lkp} illustrates the results for the cross
section as a function of LKP DM mass.  In the figure, we take $\Delta
m=50$, $100$, $150$ and $200~{\rm GeV}$ from top to bottom. The SM
Higgs boson mass is set to be 120 GeV.  It is found that the SI cross
section is enhanced as the mass difference $\Delta m$ becomes small,
while the dependence of $\Delta m$ on the SI cross section is weaker
compared to the Wino-like neutralino DM case. This is due to the SM
Higgs boson exchange contribution, which does not depend on $\Delta
m$.  It becomes sizable in the effective coupling when the other
contributions are suppressed with large $\Delta m$.

%%%%%%%%%%%%%%%%%%FIGURE%%%%%%%%%%%%%%%%%%%
\begin{figure}[t]
 \begin{center}
  \includegraphics[height=8cm]{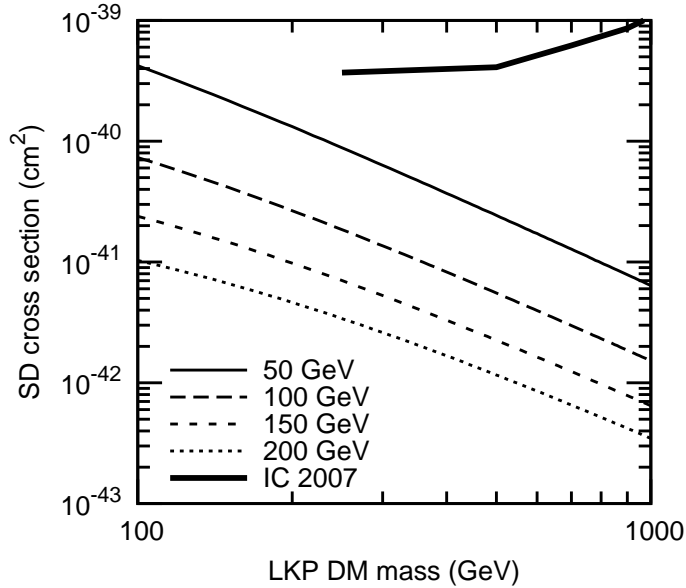}
  \caption{SD scattering cross section of LKP DM with a proton as a
    function of DM mass. Parameters are taken the same as in
    Fig.~\ref{fig:lkp}. Bold line shows IceCube bound for LKP DM
  given in Ref.~\cite{Abbasi:2009vg}.} 
\label{fig:lkpsd}
 \end{center}
\end{figure}
%%%%%%%%%%%%%%%%%%%%%%%%%%%%%%%%%%%
We show the result for the SD cross section in
Fig.~\ref{fig:lkpsd}. The parameters are taken the same as in
Fig.~\ref{fig:lkp}.  As in the previous pure Wino scenario, the
small mass difference leads to the large SD scattering cross
section. This fact indicates that the experiments
which observe the SD cross section would provide property of the dark
matter.

%%%%%%%%%%%%%%%%%%%%%%%%%%%%%%%%%%%
\section{Conclusion}
%%%%%%%%%%%%%%%%%%%%%%%%%%%%%%%%%%%%%
In this Letter we have investigated direct detection of WIMP dark
matter degenerate with new colored particles in mass. We considered
the scenario where WIMP dark matter interacts with new colored
particles and quarks. As typical examples, we studied the Wino DM in
the MSSM and the LKP DM in the MUED. Then we found that the scattering
cross section of the DM with nucleon reaches the current bound when
mass difference of the colored particle and DM is less than about $ 100~{\rm
  GeV}$ with the DM mass below $1~{\rm TeV}$.  This result shows that
the current and future direct detection experiments might shed light
on the nature of dark matter and new colored particles when their
masses are degenerate.

%%%%%%%%%%%%%%%%%%%%%%%%%%%%%%%%%%%%
\section*{Acknowledgments}
%%%%%%%%%%%%%%%%%%%%%%%%%%%%%%%%%%%%
This Letter is supported by Grant-in-Aid for Scientific research from
the Ministry of Education, Science, Sports, and Culture (MEXT), Japan,
No. 20244037, No. 20540252, No. 22244021 and No.23104011 (J.H.), and
also by World Premier International Research Center Initiative (WPI
Initiative), MEXT, Japan.  This Letter was also supported in part by the
U.S. Department of Energy under contract No. DE-FG02-92ER40701, and by
the Gordon and Betty Moore Foundation (K.I.).

%%%%%%%%%%%%%%%%%%%%%%%%%%%%%%%%%%%%
{}
%%%%%%%%%%%%%%%%%%%%%%%%%%%%%%%%%%%%

\end{document}